\documentclass[lettersize,journal]{IEEEtran}
\usepackage{subfigure}
\usepackage{amsmath,amsfonts}
\usepackage{algorithmicx}
\usepackage{algpseudocode}
\usepackage{algorithm}
\usepackage{array}
\usepackage[caption=false,font=normalsize,labelfont=sf,textfont=sf]{subfig}
\usepackage{textcomp}
\usepackage{stfloats}
\usepackage{url}
\usepackage{verbatim}
\usepackage{graphicx}
\usepackage{cite}
\usepackage{amsthm}  
\usepackage{color}
\usepackage{subcaption}
\theoremstyle{definition}
\newtheorem{definition}{Definition}
\newtheorem{assumption}{Assumption}

\theoremstyle{plain}

\usepackage{tikz}
\usepackage{tabularx}
\usepackage[table,xcdraw]{xcolor}

\begin{document}

\title{A Physics-Informed Fixed Skyroad Model for Continous UAS Traffic Management (C-UTM)}

\author{Muhammad Junayed Hasan Zahed and Hossein Rastgoftar~\IEEEmembership{Member,~IEEE,}
\thanks{The authors are with the Aerospace and Mechanical Engineering Department, the University of Arizona, Tucson,  AZ, USA 85719 email: \{mjhz, hrastgoftar\}@arizona.edu}
}



\maketitle

\begin{abstract}
Unlike traditional multi-agent coordination frameworks, which assume a fixed number of agents, UAS traffic management (UTM) requires a platform that enables Uncrewed Aerial Systems (UAS) to freely enter or exit constrained low-altitude airspace. Consequently, the number of UAS operating in a given region is time-varying, with vehicles dynamically joining or leaving even in dense, obstacle-laden environments. The primary goal of this paper is to develop a computationally efficient management system that maximizes airspace usability while ensuring safety and efficiency. To achieve this, we first introduce physics-informed methods to structure fixed skyroads across multiple altitude layers of urban airspace, with the directionality of each skyroad designed to guarantee full reachability. We then present a novel Continuous UTM (C-UTM) framework that optimally allocates skyroads to UAS requests while accounting for the time-varying capacity of the airspace. Collectively, the proposed model addresses the key challenges of low-altitude UTM by providing a scalable, safe, and efficient solution for urban airspace usability.
\end{abstract}



\section{Introduction}
Traffic management in urban areas has posed a persistent societal challenge, with considerable economic losses, environmental degradation, and reduced quality of life. Ground transportation infrastructures are constrained by physical limitations that inhibit capacity expansion, and despite the deployment of advanced traffic control and management strategies, congestion remains largely unresolved.

The emergence of uncrewed aerial systems (UAS), supported by advances in artificial intelligence, offers a paradigm shift in urban air mobility \cite{app13020755, LINGRUI2024187}. UAS can provide functions traditionally performed by ground vehicles, such as transporting goods \cite{7822209}, medical supplies \cite{THIELS2015104,Eichleay500}, and even passengers \cite{10.1063/5.0241897}. Yet, the absence of a scalable infrastructure to authorize and regulate low-altitude UAS operations constitutes a critical barrier to their widespread adoption.

To address this gap, this paper introduces a physics-informed traffic management infrastructure for urban low-altitude airspace. The framework is based on fixed skyroads, which are directionally assigned aerial corridors generated in multiple layers within the low-altitude urban environment. These skyroads enable safe, coordinated, and scalable operation of large UAS populations. By embedding them within a unified traffic management system, we establish a foundation for {\color{black}safety}, efficiency, resilience, and fairness in future urban air mobility.

\subsection{Related Work}\label{Related Work}

The integration of UAS into low-altitude airspace has necessitated the development of robust UTM systems to ensure safe and efficient operations. Key areas of research in this domain include collision avoidance and optimal path planning algorithms, fixed airspace corridors, and autonomous supervisory control mechanisms.

Path planning is essential for UAS navigation, aiming to compute collision-free and efficient routes. The $A^*$ search algorithm is widely recognized for its optimality and computational efficiency in this context \cite{8397948}. Enhancements have been proposed to the traditional $A^*$ algorithm to better accommodate the dynamic and complex environments encountered by UAS. For example, an improved $A^*$ algorithm has been developed to address the challenges of Unmanned Aerial Vehicle (UAV) path planning, focusing on refining the evaluation function and the node selection process to improve the optimality of the path and the computational speed \cite{9084806}. The cooperative $A^*$ algorithm has been designed to identify conflict-free paths by navigating space and time on a first-come, first-served principle \cite{silver2005cooperative}. However, discrete search-based path finding algorithms, like $A^*$, face challenges in scaling effectively to three dimensions or handling the finely discretized environments required for navigating complex low-altitude obstacle spaces. A common approach to accelerating these algorithms involves modifying the environment's topology to reduce the search space, albeit at the cost of solution precision specially in the context of multi-agent system. 

Various approaches have been explored for controlling multi-agent systems in UAV operations. Graph-based multi-agent reinforcement learning techniques have been developed for UAV swarm control, leveraging decentralized decision-making for coordinated navigation \cite{ZHAO2024109166}. Additionally, multi-agent system performance has been evaluated using a Kalman filter-based approach integrated with graph theory \cite{WANG2021106628}. Ensuring safety in multi-agent systems, researchers have characterized continuum deformation strategies by enforcing inter-agent collision avoidance and containment of follower agents \cite{RASTGOFTAR2021106843}. Furthermore, a hybrid AI-driven 4D trajectory management system has been proposed for urban air mobility (UAM) to optimize flight scheduling and airspace utilization \cite{XIE2024109422}. Fixed air corridor assignments have been employed in the context of UAM and UTM. Researchers proposed UTM with manually designed fixed airspace corridors \cite{pathiyil2016enabling}, multi-layered air corridors \cite{s21227536} and compared the effects of different corridor topologies on safety and efficiency \cite{sunil:hal-01168662}. However, the proposed corridors assume navigation through unrestricted airspace well above buildings and terrain, which consequently limits access for other aircraft, including passenger-carrying Advanced Air Mobility (AAM) vehicles. These vehicles are likely to have higher-priority access to the free urban airspace above the terrain compared to smaller UAS. Previously, researchers developed lane-based approach to UTM \cite{sacharny2022lane, sacharny2020efficient, henderson2019efficient, sacharny2019lane, sacharny2020faa, sacharny2020large, sacharny2020dddas, henderson2023multi, emadi2022physics} where airway corridors are designed based on ground road networks, with flights progressing through a sequence of lane transitions from take-off to landing. An alternative UTM concept presented in \cite{10334303}, discusses the concept of finite-state fixed airspace corridors by using Eulerian continuum mechanics as referenced in \cite{8907366, uppaluru2022resilient, rastgoftar2020fault}, to partition airspace into navigable channels that safely wrap around unplanned zones, such as buildings and restricted areas. Collision prevention in this framework \cite{10334303} is facilitated by a Markov Decision Process (MDP), which allocates UAS to corridors based on a first-come, first-served protocol. 

While these studies have contributed significantly to multi-agent system control, they do not account for the dynamic nature of UAV numbers within the system. Most existing approaches assume a fixed number of agents, limiting their applicability in scenarios where UAVs continuously enter and exit the airspace. Addressing this limitation, our work focuses on a dynamic multi-agent framework that enables adaptive UAV traffic management in urban airspace. {\color{black}In this context, we integrate an autonomous supervisory control system into UAS operations to facilitate real-time decision-making and dynamic traffic management on a first-come, first-served basis.}

\begin{figure}[t]
  \centering
  \setlength{\tabcolsep}{2pt} 
  \begin{tabular}{@{}cccc@{}}
    \includegraphics[width=0.48\textwidth]{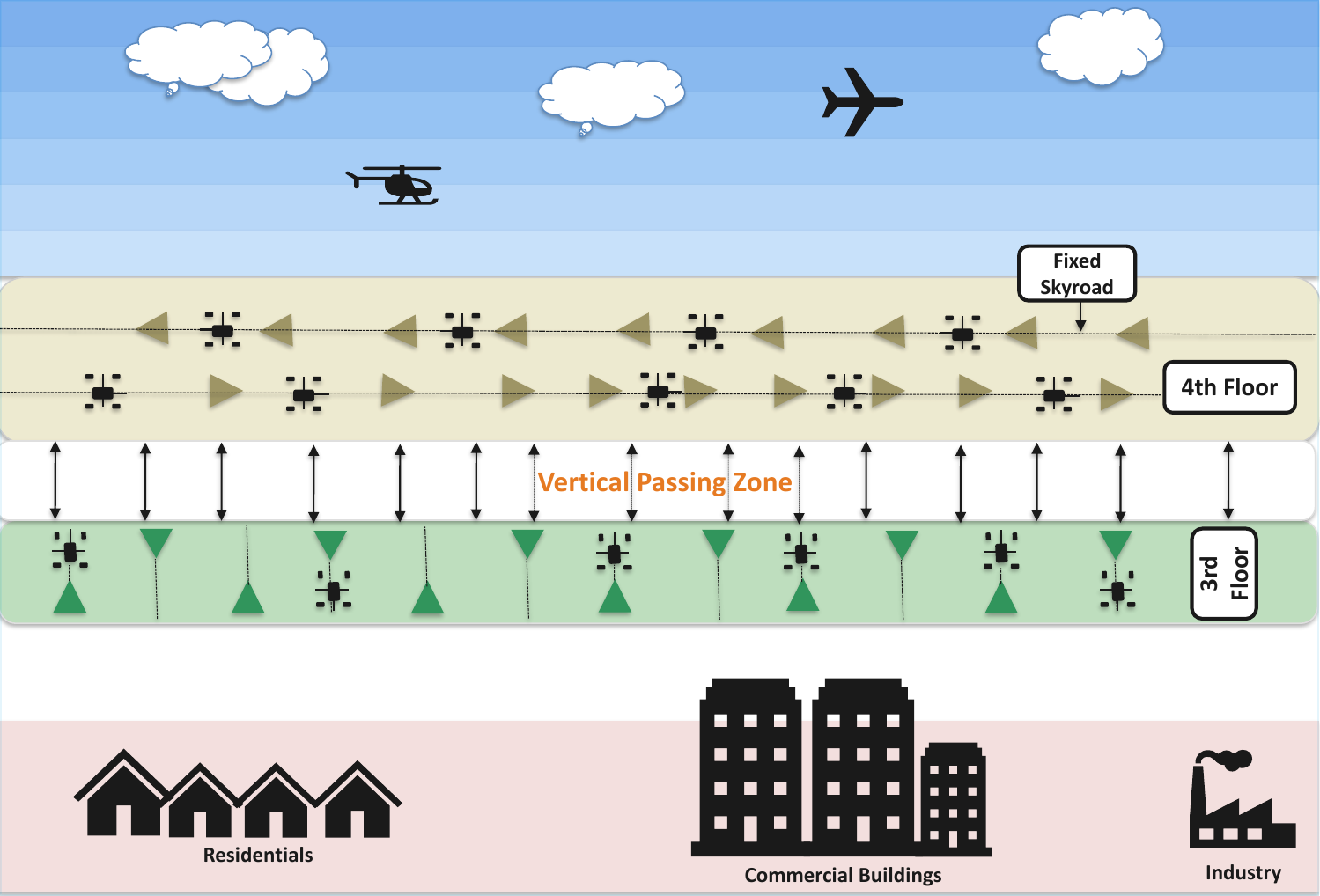}

  \end{tabular}
  \caption{\color{black}C-UTM illustrating fixed skyroads with vertical passing zone. In each layer the skyroads are multi-track and bidirectional, while skyroads on the adjacent layers are orthogonal to each other.}
  \label{Illustration}
\end{figure}

UAM Concept of Operations (ConOps) v2.0 anticipates an evolution from simple, one way UAM corridors to higher capacity constructs using Automated Flight Rules (AFR), passing zones, and multi track internal structure as UAM operations tempo rises \cite{conops_v2}. C-UTM's vertically stacked, orthogonal floors and fixed skyroads operationalize this evolution. Fig.\ref{Illustration}  depicts a vertical passing zone between floors directly analogous to ConOps' vertical passing concepts allowing orderly overtakes without violating floor directionality. Further, ConOps frames UAM corridors as cooperative areas within the complementary Air Traffic Services (ATS)/ Extensible Traffic Management (xTM) service environments, governed by FAA approved Cooperative Operating Practices (COPs), federated information exchange, and demand capacity balancing (DCB). In sum, the fixed skyroad and C-UTM framework collectively furnish the operational mechanics,  structure, separation logic and allocative fairness needed for ConOps' corridor evolution from low tempo beginnings to dense, automated and federated UAM networks.

\subsection{Contributions}\label{Contributions and Problem Statement}
This paper aims to enable continuous operation of UAS in the low-altitude airspace of urban environments, which is constrained by buildings of varying sizes, heights, and shapes, as well as other stationary objects. To achieve this goal, we propose defining fixed skyroads across multiple horizontal floors at different altitudes, where the nominal motion directions of adjacent floors are chosen to be mutually orthogonal to ensure full reachability of the airspace. Within each floor, fixed skyroads are generated using an irrotational fluid-flow model, characterized by potential and stream functions that both satisfy the Laplace partial differential equation (PDE).

For the proposed UTM framework, the streamlines determine the geometry (boundaries) of the skyroads and must be constructed such that the following conditions are satisfied:

\noindent \textbf{Condition 1:} The keep-out zones—defined by the boundaries of buildings and stationary objects of arbitrary size and geometry—are fully enclosed by the streamlines without intersecting the keep-out regions. This requirement must hold regardless of the random size, geometry, and number of keep-out zones produced by cutting through buildings and stationary obstacles.

\noindent \textbf{Condition 2:} The skyroads are generated such that their orientation aligns with the prescribed nominal direction of the corresponding floor. This ensures that all skyroads within a floor are approximately uniform and follow the desired flow direction.

The \textbf{preliminary }contribution of this paper is the development of a generic solution for specifying skyroads within each floor of the airspace while ensuring that Conditions 1 and 2 are satisfied. To this end, we propose a structured approach for determining the external boundary conditions as well as the conditions along the boundaries of the keep-out zones, such that the stream functions—obtained by solving the Laplace PDE—simultaneously satisfy both constraints. By defining skyroads within each floor, we assign a unique motion direction to every skyroad, which must be strictly followed by any UAS  operating within it. The directionality of the skyroads is also determined to ensure that the entire airspace remains reachable. To enforce this directionality, we spatially discretize each skyroad into a finite number of segments defined by the set $\mathcal{V}$, and employ the graph $\mathcal{G}_{\text{glob}}\left(\mathcal{V},\mathcal{E}\right)$ to structure transitions across the skyroads, where the edge set $\mathcal{E}\subset \mathcal{V}\times \mathcal{V}$ defines all possible transitions.

While ConOps v2.0 envisions bidirectional, multi-track corridors within a layer, augmented by both vertical and lateral passing zones \cite{conops_v2}, our skyroad generation model constructs fixed airways that are multi-track and bidirectional within every layer, while adjacent layers are orthogonal, furnishing additional, geometry-driven reachability across the urban airspace. To streamline operations and preserve strong invariants on flow direction, we employ vertical passing zones only, eliminating lateral merges/diverges within a layer.

\begin{figure}[h]
    \centering
    \includegraphics[width=\linewidth]{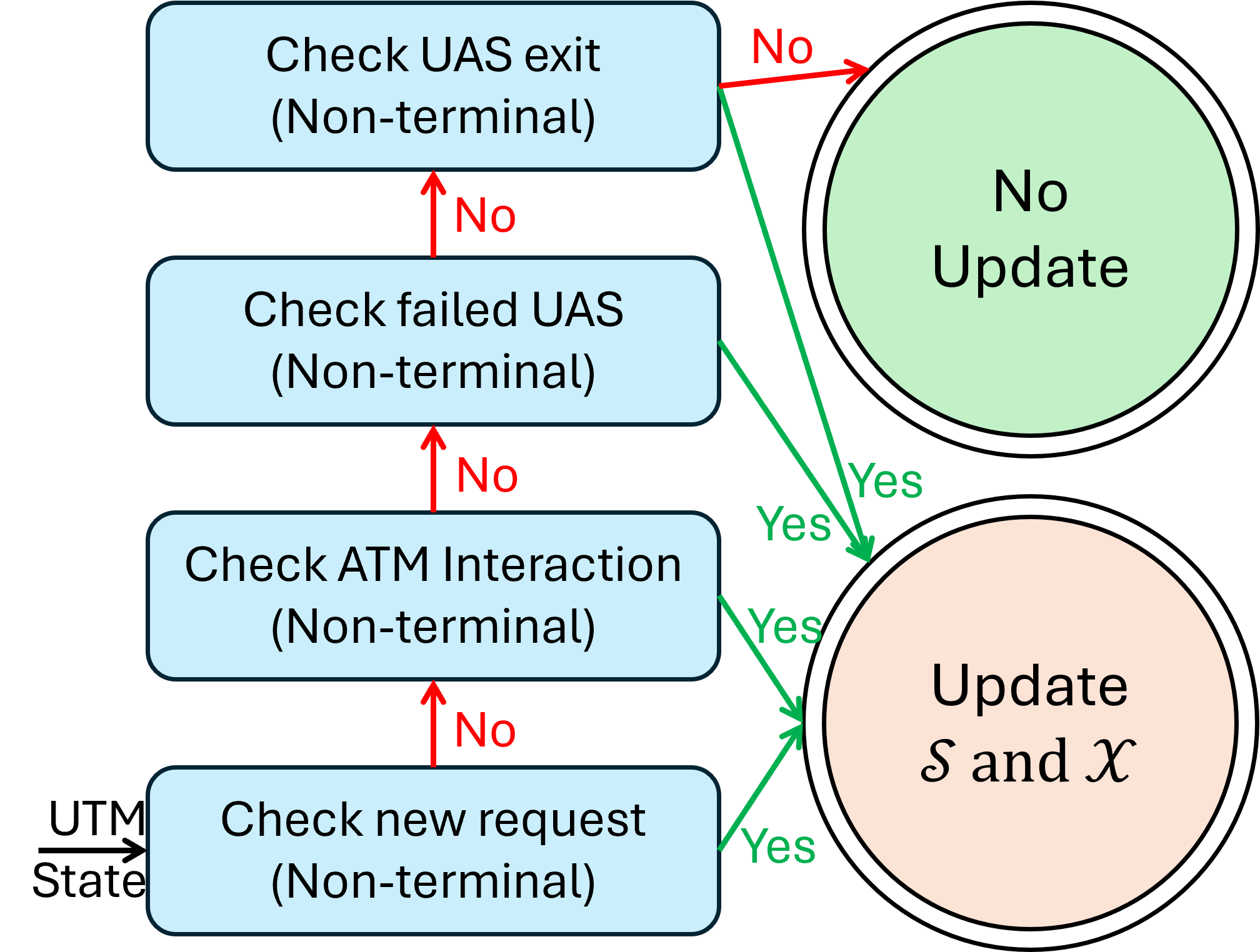}
    \caption{The state machine used to update accessible node $\mathcal{W}$ and edge set $\mathcal{X}\subset \mathcal{W}\times \mathcal{W}$ specifying transition over conflict-free airspace.} \label{StateMachine} 
\end{figure}

The \textbf{primary} contribution of this paper is the development of a continuous UAS traffic management (C-UTM) system that maximizes the usability of the airspace while enforcing UAS to utilize the skyroads and follow the desired motion direction assigned to each skyroad. The C-UTM is decomposed into \textit{supervisory control} and \textit{skyroad allocation} modules. The supervisory control module implements the state machine shown in Fig.~\ref{StateMachine} to consistently divide the skyroad segments, defined by $\mathcal{V}$, into accessible segments $\mathcal{W}$ and allocated segments $\bar{\mathcal{W}}$. Transitions over the accessible skyroads are then structured by updating the graph $\mathcal{G}\left(\mathcal{W},\mathcal{X}\right)$, where the edge set $\mathcal{X}\subset \mathcal{W}\times \mathcal{W}$ defines the allowable transitions across accessible skyroad segments. Given the graph $\mathcal{G}\left(\mathcal{W},\mathcal{X}\right)$, the skyroad allocation module applies a search method to allocate air corridors to new requests, prioritizing the submission time in order to respond fairly to UAS requests.

\subsection{Outline}
The related work presented in Section \ref{Related Work}, and the contributions and problem statement presented in Section \ref{Contributions and Problem Statement}, are followed by the Problem Statement presented in Section \ref{Problem Statement}. Our methodology is decomposed into skyroad generation and skyroad operation and presented in Sections \ref{Skyroad Generation} and \ref{Skyroad Operation}, respectively. The simulation results are presented in Section \ref{Results} and followed by Conclusion in Section \ref{Conclusion}.

\section{Problem Statement}\label{Problem Statement}

The concept of this paper is decomposed into \textbf{skyroad generation} and \textbf{skyroad operation}. For skyroad generation, UAS transportation in the urban airspace is represented by $\mathcal{G}_{glob}\left(\mathcal{V},\mathcal{E}\right)$, where $\mathcal{V}$ denotes all skyroad segments, and $\mathcal{E}\subset \mathcal{V}\times \mathcal{V}$ defines the authorized transitions between these segments. Section \ref{Skyroad Generation} explains how the principles of fluid mechanics are applied to define skyroads that are uniformly specified across each floor along the desired direction, and how they are discretized to ensure full reachability of the airspace.

Skyroad operation encompasses \textbf{supervisory control} and \textbf{skyroad allocation} modules. The supervisory control employs the state machine shown in Fig. \ref{StateMachine} to algorithmically update the set of allocated skyroad segments, denoted by $\bar{\mathcal{W}}\subset \mathcal{V}$; define the set of accessible states as $\mathcal{W}=\mathcal{V}\setminus \bar{\mathcal{W}}$; and restructure transitions over the accessible skyroad segments by updating $\mathcal{G}\left(\mathcal{W}, \mathcal{X}\right)$. The skyroad allocation module then applies the $A^*$ search on the most recent graph $\mathcal{G}\left(\mathcal{W}, \mathcal{X}\right)$ to allocate skyroads to new UAS requesting access to the airspace.

\begin{figure}[h!]
    \centering
    \includegraphics[width=90mm, height = 70mm]{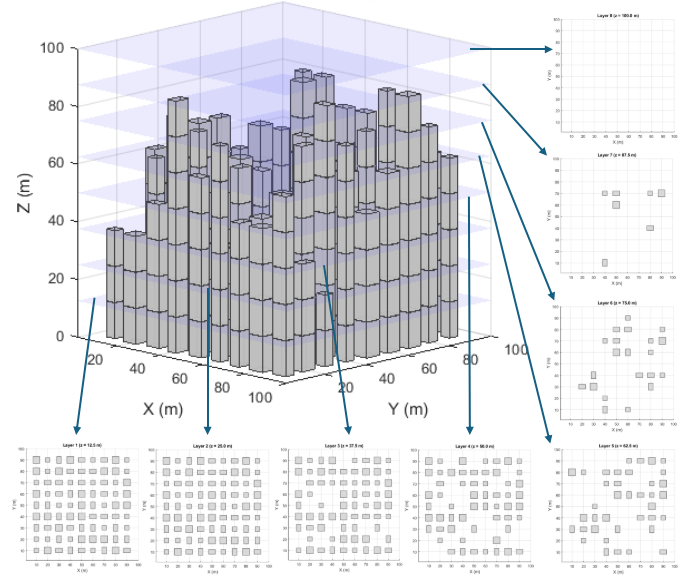}
    \caption{{A virtual urban landscape featuring numerous buildings of varying lengths, widths, and heights. Eight layers are spaced 12.5 meters apart. Zoomed in view of each 2D layer in XY plane is directed by arrow marks.}} \label{landscape}
\end{figure}

\section{Approach}\label{Approach}
We divide finite airspace into multiple floors $\mathcal{P}_1$  through $\mathcal{P}_{n_z}$, where $\mathcal{P}_h$ is a horizontal floor at elevation $z_h\in \mathbb{R}_+$,  ($h = 1,\cdots ,n_z$), and $x$ and $y$ are used to specify position in every floor. Floor $\mathcal{P}_h$ is decomposed into keep-in subspace $\mathcal{M}_h$ and keep-out subspace $\mathcal{R}_h$ ($\mathcal{P}_h  =\mathcal{M}_h \cup \mathcal{R}_h$), where keep-out subspace(s), or zone(s) enclose  buildings, structures, and restricted no-fly regions. For better clarification, consider the schematic of an urban airspace shown in Fig.  \ref{landscape}, where it is cut by $n_z=8$ floors with the keep-in and keep-out zones shown on the right and bottom sides respectively.

\subsection{Skyroad Generation}
\label{Skyroad Generation}

\begin{figure}[h]
    \centering
    \includegraphics[width=80mm, height = 65mm]{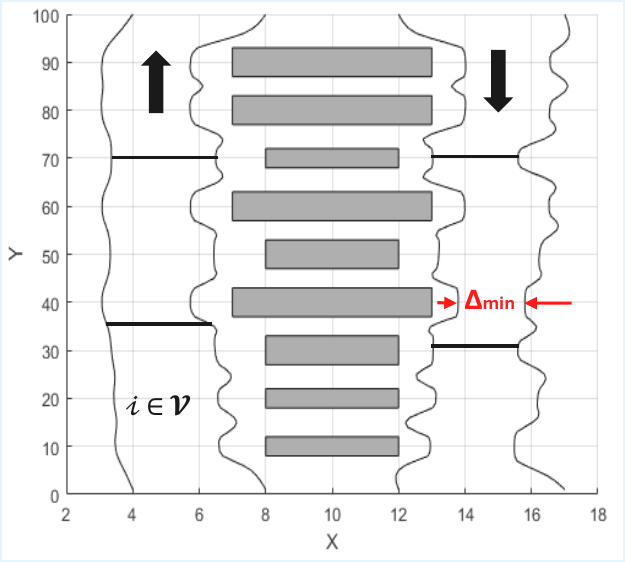}
    \caption{{Skyroads along certain directions sandwitching the keep-out zones {\color{black} Please fix the picture. Just two sky roads on the left and two skyroads on the right. Each skyroad needs to be segmented and illustrated. Set $\mathcal{V}$ defines skyroad segments not nodes.}}} \label{Streamline_Directions}
    
\end{figure}

We desire to define skyroads  in the  keep-in subspace of every floor so that the following objectives are achieved:
\begin{enumerate}
    \item Every keep-out zone is safely wrapped by a streamline, which can be defined as a skyroad.
    \item Skryroads are all directed along a certain direction in every floor while keep-out zones are sandwitched (see Fig. \ref{Streamline_Directions}).  
\end{enumerate}

\subsubsection{Foundations}

We treat UAS coordination as an ideal-fluid flow to achieve the above objectives. Therefore, we can define potential field $\phi^h\left(x,y\right)$ and stream field $\psi^h\left(x,y\right)$ over the floor $h$, where both stream and potential fields satisfy the Laplace equation: 
\begin{subequations}
    \begin{equation}
        {\partial^2\phi^h\left(x,y\right)\over \partial x^2}+{\partial^2\phi^h\left(x,y\right)\over \partial y^2}=0,\qquad h=1,\cdots,n_z,
    \end{equation}
    \begin{equation}
        {\partial^2\psi^h\left(x,y\right)\over \partial x^2}+{\partial^2\psi^h\left(x,y\right)\over \partial y^2}=0,\qquad h=1,\cdots,n_z.
    \end{equation}
\end{subequations}
Note that we do not need to solve for $\phi^h\left(x,y\right)$ to specify the skyroads as the skyroads lie along the streamlines in every floor $h$. Indeed, $\phi^h$ can determine desired speeds along the stream lines in every floor $h$ which are not needed for the high-level planning and management problems we consider in this paper. Therefore, we only need to solve $\psi^h\left(x,y\right)$ for the arbitrary distribution and size of the keep-out zones. We use the Finite Diference method to obtain $\psi^h\left(x,y\right)$ in every floor $h$ where we apply the boundary conditions presented in Sections \ref{External Boundary Conditions} and \ref{Conditions on the Boundary of the Keep-Out Zones}, respectively.

\subsubsection{Skyroad Structure}\label{External Boundary Conditions}
\begin{figure}[h]
    \centering
    \includegraphics[width=0.8\linewidth]{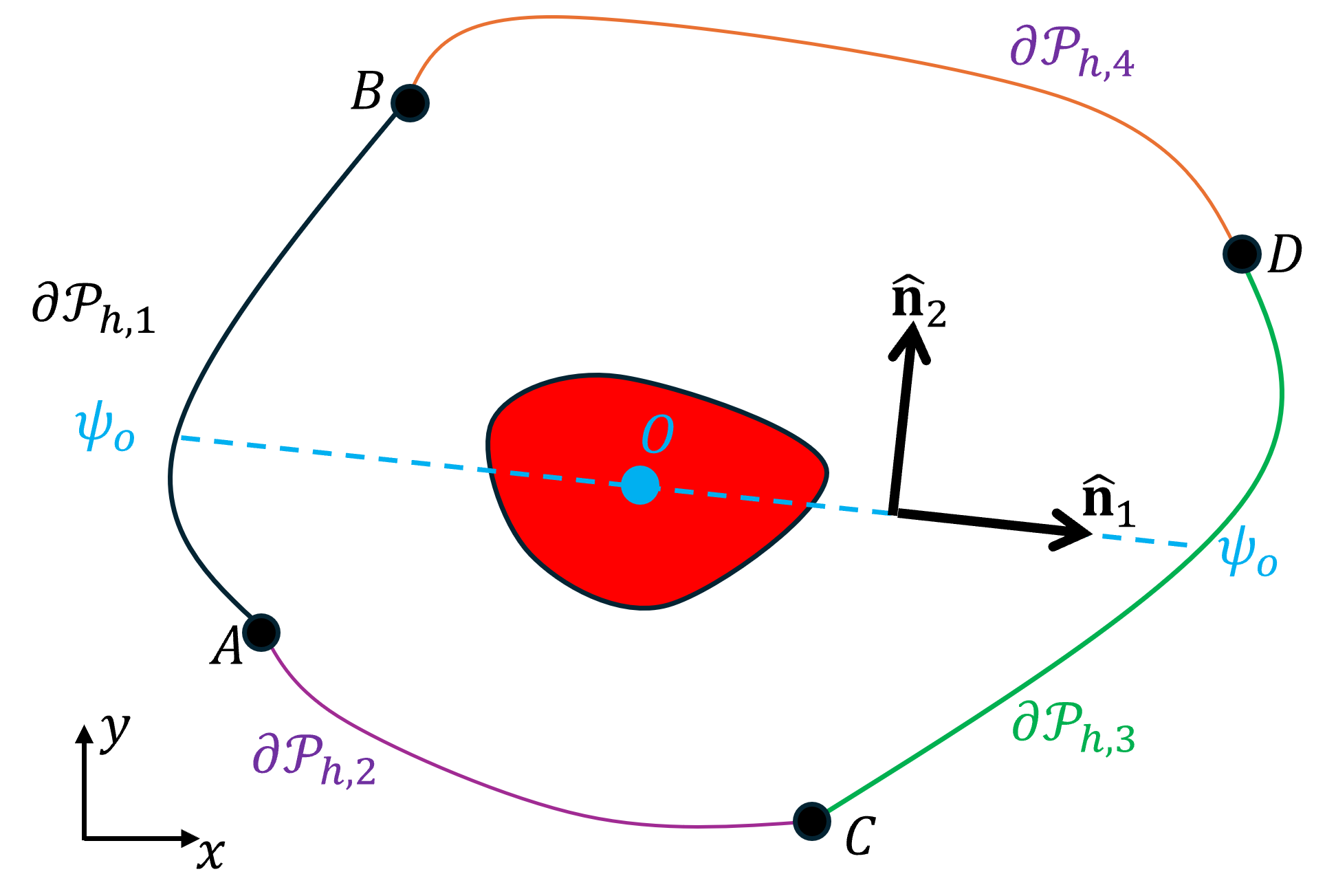}
    \caption{The proposed segmentation of the boundary of floor $h$} \label{BC} 
\end{figure}

The boundary of $\mathcal{P}_h $ is divided into four segments $\partial \mathcal{P}_{h,1}$, $\partial \mathcal{P}_{h,2}$, $\partial \mathcal{P}_{h,3}$, $\partial \mathcal{P}_{h,4}$ (see Fig. \ref{BC}). The stream value along boundaries $\partial \mathcal{P}_{h,2}$ and $\partial \mathcal{P}_{h,4}$ are constant and equal to $\psi_{min}$ and $\psi_{max}$, respectively. To obtain the stream values along boundaries $\partial \mathcal{P}_{h,1}$ and $\partial \mathcal{P}_{h,3}$, we define the following terms:

\begin{definition}
    The intersection point of boundaries $\partial \mathcal{P}_{h,1}$ and $\partial \mathcal{P}_{h,2}$ are marked as  point $A$ where $\mathbf{r}_A$ denotes the position of $A$ with respect to an inertial coordinate system.
\end{definition}
\begin{definition}
    The intersection point of boundaries $\partial \mathcal{P}_{h,1}$ and $\partial \mathcal{P}_{h,4}$ are marked as  point $B$ where $\mathbf{r}_B$ denotes the position of $B$ with respect to an inertial coordinate system.
\end{definition}
\begin{definition}
    The intersection point of boundaries $\partial \mathcal{P}_{h,2}$ and $\partial \mathcal{P}_{h,3}$ are marked as  point $C$ where $\mathbf{r}_C$ denotes the position of $B$ with respect to an inertial coordinate system.
\end{definition}
\begin{definition}
    The intersection point of boundaries $\partial \mathcal{P}_{h,3}$ and $\partial \mathcal{P}_{h,4}$ are marked as  point $D$ where $\mathbf{r}_D$ denotes the position of $D$ with respect to an inertial coordinate system.
\end{definition}

\begin{definition}
    We define $\hat{\mathbf{n}}_1$ as the desired nominal orientation of the streamlines (skyroads) in floor $h$.
\end{definition}
\begin{definition}
    We define $\hat{\mathbf{n}}_2$ as a vector normal to $\hat{\mathbf{n}}_1$  obtained by counterclockwise rotation along the positive $z$ axis.
\end{definition}

Given the above definitions, the stream values along boundaries $\partial \mathcal{P}_{h,1}$ and $\partial \mathcal{P}_{h,3}$ are defined as follows: 
\begin{subequations}
\begin{equation}
\begin{aligned}
    \psi^h(\mathbf{r})=\psi_{min}+{\left(\mathbf{r}-\mathbf{r}_A\right)\cdot\hat{\mathbf{n}}_2\over \left(\mathbf{r}_B-\mathbf{r}_A\right)\cdot\hat{\mathbf{n}}_2}\left(\psi_{max}-\psi_{min}\right),\\
    \qquad \forall \mathbf{r}\in \partial \mathcal{P}_{h,1},~h=1,\cdots,n_z
\end{aligned}
\end{equation}   
\begin{equation}
\begin{aligned}
    \psi^h(\mathbf{r})=\psi_{min}+{\left(\mathbf{r}-\mathbf{r}_C\right)\cdot\hat{\mathbf{n}}_2\over \left(\mathbf{r}_D-\mathbf{r}_C\right)\cdot\hat{\mathbf{n}}_2}\left(\psi_{max}-\psi_{min}\right),\\
    \qquad \forall \mathbf{r}\in \partial \mathcal{P}_{h,3},~h=1,\cdots,n_z
\end{aligned}
\end{equation} 
\end{subequations}

\subsubsection{Conditions on the Boundary of the Keep-Out Zones}\label{Conditions on the Boundary of the Keep-Out Zones}
The stream value is constant everywhere along the external boundary of an obstacle. Let center point of a keep-out zone is marked as the point $O$  where $\mathbf{r}_O$ denotes the position of $O$ with respect to an inertial coordinate system. To obtain the stream value on the boundary of a keep-out zone, we make the following assumption:
\begin{assumption}\label{boundaryassumption}
    We divide boundaries of floor $h$ such that vectors $\mathbf{r}_C-\mathbf{r}_A$ and $\mathbf{r}_D-\mathbf{r}_B$ are both parallel to unit vector $\hat{\mathbf{n}}_1$. 
\end{assumption}

When Assumption \ref{boundaryassumption} is satisfied, a parallel line to $\hat{\mathbf{n}}_1$ crossing boundaries  $\partial \mathcal{P}_{h,1}$ and $\partial \mathcal{P}_{h,3}$ at  the same stream value that is denoted $\psi_o$ (see the line colored in cyan in Fig. \ref{BC}). Then, stream value along the external boundary of this keep-out zone is constant and equal to $\psi_o$.

\subsubsection{Safety} 
The skyroad  boundaries can safely wrap all the buildings and stationary infrastructures. However, we need  to assign the minimum bandwidth  $\Delta_{min}$ for each skyroad based on the UAS mission operational condition and application (see Fig. {\color{black}\ref{Streamline_Directions}}). Therefore, streamlines assigned by solving the proposed Laplace PDE are constellated such that this safety requirement is met. As the result, our proposed physics-informed model can leverage heterogeneous UAS mission and applications while ensuring safety and collision avoidance with existing buildings and stationary objects. We note that inter-UAS collision avoidance can be properly addressed through the proposed UAS operation model discussed in Section \ref{Skyroad Operation}.
\subsubsection{Stracturing the Transportation Network}
\begin{figure*}[t]
  \centering
  \setlength{\tabcolsep}{2pt} 
  \begin{tabular}{@{}cccc@{}}
    \includegraphics[width=.23\textwidth]{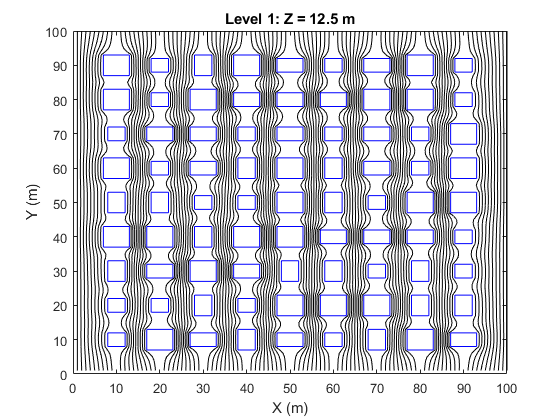} &
    \includegraphics[width=.23\textwidth]{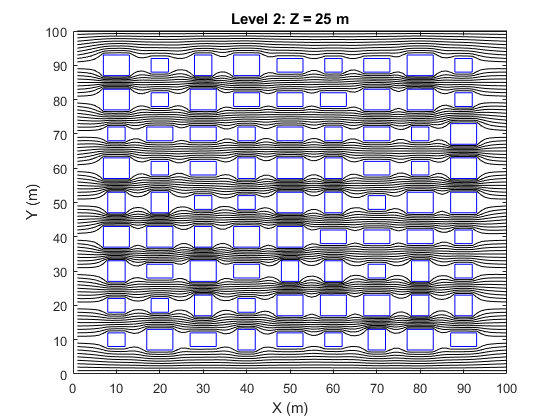} &
    \includegraphics[width=.23\textwidth]{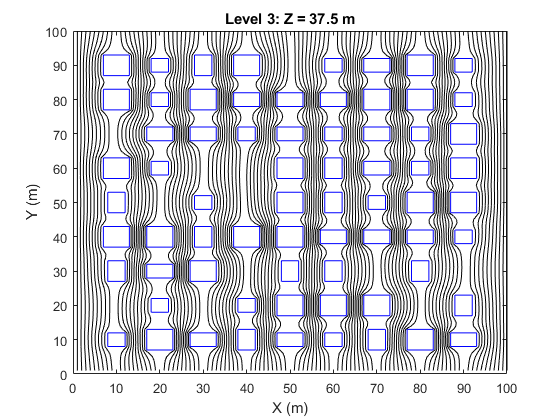} &
    \includegraphics[width=.23\textwidth]{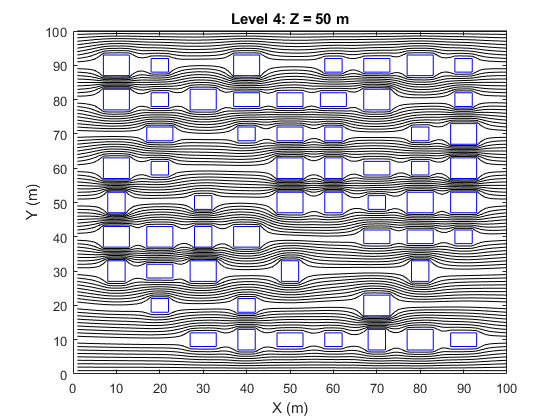} \\
    \includegraphics[width=.23\textwidth]{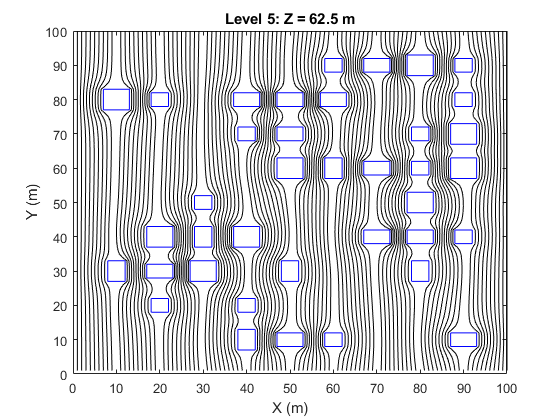} &
    \includegraphics[width=.23\textwidth]{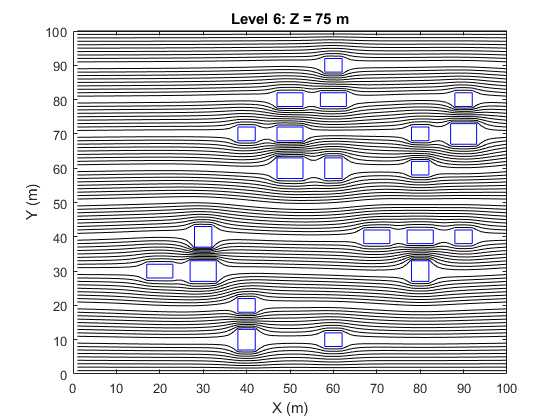} &
    \includegraphics[width=.23\textwidth]{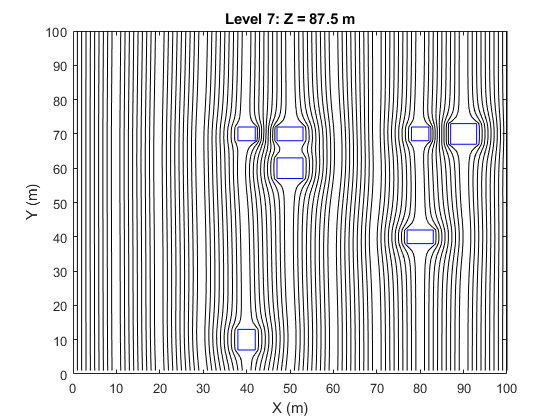} &
    \includegraphics[width=.23\textwidth]{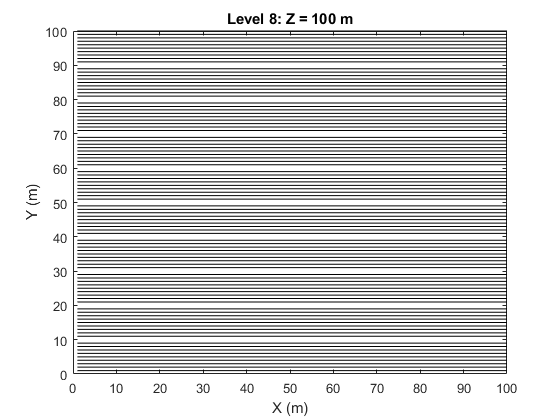}
  \end{tabular}
  \caption{Fixed skyroads across Levels 1–8.}
  \label{Fixed_Corridors}
\end{figure*}

\begin{figure}[h!]
    \centering
    \includegraphics[width=85mm, height = 65mm]{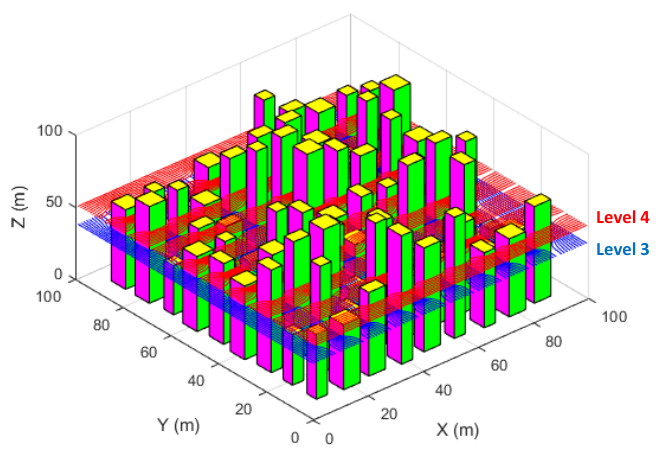}
    \caption{{Skyroads on the 3rd and 4th floor wrapping the buildings.}} \label{3rd_4th_wrapped}
    
\end{figure}

\begin{figure}[t]
  \centering
  \begin{minipage}[b]{.49\columnwidth}
    \centering
    \includegraphics[width=\linewidth]{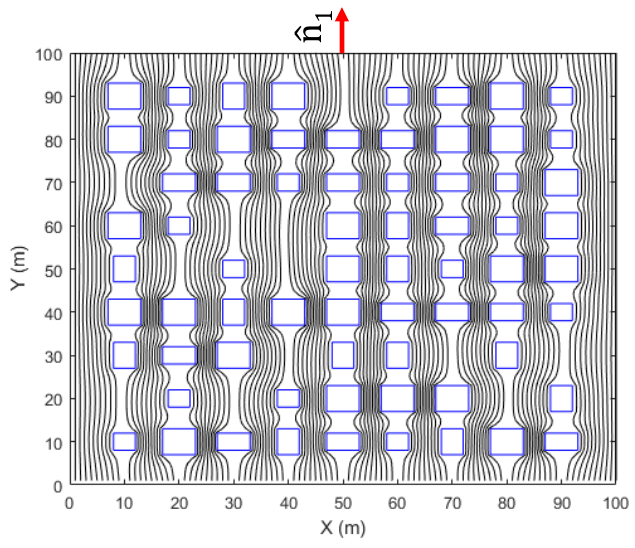}
  \end{minipage}\hfill
  \begin{minipage}[b]{.49\columnwidth}
    \centering
    \includegraphics[width=\linewidth]{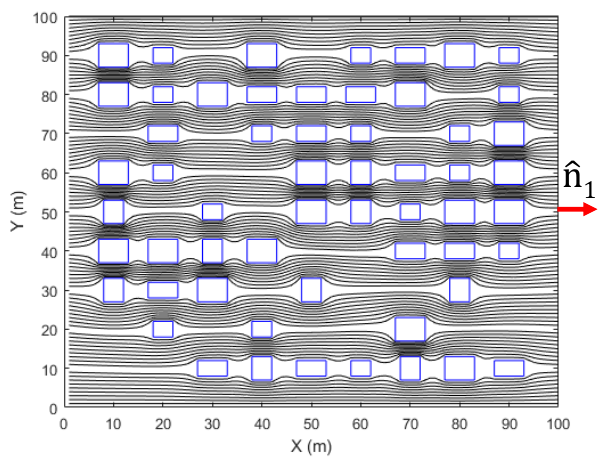}
  \end{minipage}
  \caption{Nominal motion directions on the 3rd and 4th floors are perpendicular.}
  \label{Perpendicular}
\end{figure}

We use graph $\mathcal{G}_{glob}\left(\mathcal{V},\mathcal{E}\right)$ to structure transitions over the skyroad of desire. We divide each styroad to mutiple segments and set $\mathcal{V}$ defines all possible skyroad segements (see Fig. \ref{Streamline_Directions}). Set $\mathcal{E}$ specifying transitions over the skyroad segement set $\mathcal{V}$ is determined by applying the following rules:
\begin{enumerate}
    \item Motion along each skyroad is unidirectional (see Fig. \ref{Streamline_Directions}).
    \item Two adjacent skyroads in the same floor authorize the opposite motion directions (Fig. \ref{Fixed_Corridors}).
    \item Traversal motion is not authorized in every floor $h$. This rule implies that UAS cannot change the skyroad while it has not change the floor. 
    \item Nominal motion directions in every two adjacent floors are perpendicular to each other (see Fig. \ref{3rd_4th_wrapped} and \ref{Perpendicular}).
\end{enumerate}
To ensure reachability across the airspace, nominal motion directions in every two adjacent floors are perpendicular to each other.

We note that the set $\mathcal{V}$ can be defined under both \emph{normal} and \emph{anomalous} situations.
Under normal situations, we assume that the skyroad segments defined by $\mathcal{V}$ fully cover the keep-in zones on every floor.
Under anomalous situations—such as interactions between UTM and ATM, or the failure of a UAS in the airspace that must be properly contained—the set $\mathcal{V}$ must be updated so that the restricted regions, allocated either to ATM operations or to the safety recovery of a failed UAS, are excluded from accessibility by other UAS. The latter case can be properly addressed through the skyroad operation presented in Algorithm \ref{alg2}

\subsection{Skyroad Operation}\label{Skyroad Operation}

We propose a skyroad operation model that enables a large number of UAS to simultaneously utilize the airspace by moving along allocated skyroad corridors, while ensuring safe entry, exit, and collision avoidance with buildings and other obstacles. This model is decomposed into supervisory control and skyroad allocation and presented in Sections \ref{Supervisory Control} and \ref{Skyroad Allocation}, respectively.

\subsubsection{Supervisory Control}\label{Supervisory Control}
The skyroad operation problem is inherently computationally intensive due to the challenges of large-scale coordination, the time-varying capacity of the airspace, and the non-concurrent nature of access requests for entering and exiting the system. To address this challenge, we propose a first-come, first-served (FCFS) strategy that assigns an air corridor along the accessible skyroads to a single UAS at a time, whenever a request exists, strictly based on the time of submission. 


To implement this strategy, we employ the state machine illustrated in Fig.~\ref{StateMachine}, which forms the basis of Algorithm~\ref{alg2}. This framework enables efficient updates of the accessible skyroads while enforcing the following rules:

\begin{itemize}
    \item \textbf{Rule 1:} UAS requesting to use the airspace is given a maximum time $T_{max}$. \textit{We propose to assign $T_{max}$ proportional to the number of skyroad segments along the optimal path assigned assigned to the UAS.} \label{Rule 1}
    \item \textbf{Rule 2:} Because motion along the skyroads are unidirectional, the {\color{black}skyroad segments} along the path allocated to UAS are freed up if UAS already pass those {\color{black}skyroad segments}. These already-traveled {\color{black}skyroad segments} are called \textit{completed {\color{black}skyroad segments}} and defined by set $\mathcal{F}(k)\subset \bar{\mathcal{W}}$ at discrete time $k$.  The C-UTM operation updates the available node set $\mathcal{W}$ every $N_T>1$ time steps to add the completed {\color{black}skyroad segments}  to $\mathcal{W}$, i.e $\mathcal{W}\leftarrow \mathcal{W}\cup \mathcal{F}$.
\end{itemize}
\begin{definition}
We set $\mathcal{T}_{check}(k)$ as the set aggregating the due date times for UAS to leave the airspace, where it is define as updated as follows:
\begin{equation}\label{tcheck}
    \mathcal{T}_{check}(k)=
    \begin{cases}
    \mathcal{T}_{check}(k-1)\cup\left\{k+T_{max}\right\},\\
    \hspace{1cm}\mathrm{if~a~new~air~corridor~is~allocated.}\\
    \mathcal{T}_{check}(k-1),
    \hspace{0.5cm}\mathrm{otherwise}.
    \end{cases}
\end{equation}
\end{definition}
Algorithm~\ref{alg2} applies Rules~1 and~2 by receiving the skyroad segment set $\mathcal{V}$ and the edge set $\mathcal{E}$ as input, and it begins its operation with the following initial conditions:
\[
k = 0, \quad \bar{\mathcal{W}} = \emptyset, \quad \mathcal{W} = \mathcal{V}, \quad \mathcal{T}_{check}(k) = \emptyset,
\]
which implies that all skyroads are initially available.  

A binary variable $finish \in \{0,1\}$ is introduced to determine whether the C-UTM operation continues or terminates. Specifically, the operation continues if $finish = 1$ and terminates if $finish = 0$.  

At each discrete time step $k$, Algorithm~\ref{alg2} decomposes the node set $\mathcal{V}$ into two disjoint subsets:
\[
\mathcal{V} = \mathcal{W} \cup \bar{\mathcal{W}},
\]
where $\bar{\mathcal{W}}$ denotes the set of skyroad segments already allocated to UAS operating in the designated airspace, and $\mathcal{W}$ denotes the remaining accessible skyroad segments.  

Given the edge set $\mathcal{E}$, Algorithm~\ref{alg2} also generates the set $\mathcal{X}$, which defines the authorized transitions among accessible skyroad segments. In particular,
\[
(i,j) \in \mathcal{X} \iff i \in \mathcal{W}, \; j \in \mathcal{W},
\]
meaning that node $j$ can be reached from node $i$ while remaining within the set of accessible segments.


\subsubsection{Skyroad Allocation}\label{Skyroad Allocation}
Without loss of generality, for air corridor allocation, we apply the A* search method to determine the shortest air corridors along the skyroads that need to be allocated to a UAS requesting access to the airspace. The $A^*$ planner receives the initial location $i_0\in \mathcal{W}$ and target location $i_g\in \mathcal{W}$ at discrete time $k$. The metric distance 

\begin{equation}
\begin{aligned}
    H(i) = \sqrt{(x_i-x_{i_g})^2+(y_i-y_{i_g})^2+(z_i-z_{i_g})^2},\\
    \qquad i\in\mathcal{W},\ k\in\mathbb{N}
\end{aligned}
\end{equation}
is considered as the heuristic cost, where $x_i$, $y_i$ and $z_i$ are position components of node $i\in \mathcal{W}$. By defining
\begin{equation}
    \mathcal{N}_{i,k}=\left\{j\in \mathcal{W}:\left(i,j\right)\in \mathcal{X}\right\},\qquad i\in \mathcal{W},~k\in \mathbb{N}
\end{equation}
the transition cost is defined by 
\begin{equation}
\begin{aligned}
    c\left(i,j\right)=\sqrt{\left(x_i-x_j\right)^2+\left(y_i-y_j\right)^2+\left(z_i-z_j\right)^2}\\\qquad i\in \mathcal{W},~j\in \mathcal{N}_{i,k},~k\in \mathbb{N}.
\end{aligned}
\end{equation}

\begin{algorithm*}[h!]
  \caption{C-UTM Operation}\label{alg2}
  \begin{algorithmic}[1]
        \State \textit{Get:} Node set $\mathcal{V}$ and $\mathcal{E}$;
        \State $k=0$; 
        \State $\bar{\mathcal{W}}=\emptyset$; 
        \State ${\mathcal{W}}=\mathcal{V}$;
        \State $\mathcal{T}_{check}\left(k\right)=\emptyset$;
        \State $finish=0$.
        \While{$finish = 0$}        
            \State Obtain set $\mathcal{F}$, defining the \textit{completed {\color{black}skyroad segments}} at discrete time $k$;
            \If{$\mathrm{Remainder}({k\over N_T}) = 0$ or $k\in \mathcal{T}_{check}(k)$;}   
               \State Update $\bar{\mathcal{W}}$: $\bar{\mathcal{W}}\setminus \mathcal{F}$;
               \State Update ${\mathcal{W}}$: ${\mathcal{W}}=\mathcal{V}\setminus \bar{\mathcal{W}}$;
               \State Update the transition set $\mathcal{X}$;
           \EndIf
           \If{New UAS request is submitted to use the airspace,}
                \State Allocate the optimal path to the UAS by searching over graph $\mathcal{G}\left(\mathcal{W},\mathcal{X}\right)$;
                \State Update  $\mathcal{T}_{check}\left(k\right)$ by using Eq. \eqref{tcheck};
               \State Update $\bar{\mathcal{W}}$: Add the {\color{black}skyroad segments} lying on the allocated optimal path to set $\bar{\mathcal{W}}$;
               \State Update ${\mathcal{W}}$: ${\mathcal{W}}=\mathcal{V}\setminus \bar{\mathcal{W}}$;
               \State Update the transition set $\mathcal{X}$;
           \EndIf
           \If{C-UTM operation must be stopped}
                \State $finish \gets 0$
            \Else
                \State $k \gets k+1$
                \If{an anomalous situation is detected}
                    \State Update $\mathcal{V}$ and $\mathcal{E}$ accordingly
                \EndIf
            \EndIf
        \EndWhile        
  \end{algorithmic}
\end{algorithm*}





\section{Simulation Results}\label{Results}

This study presents a simulation framework for dynamic path planning and C-UTM operations for UASs navigating a discretized, obstacle-rich 3D urban airspace, as illustrated in Figure \ref{landscape}. The simulated environment consists of a structured 91 × 91 horizontal grid, segmented into eight horizontal floors, each representing an altitude level with incremental heights of $12.5$ meters. To determine the optimal fixed skyroad for UASs, the $A^*$ search algorithm was employed, ensuring efficient and obstacle-free navigation to the designated target. The C-UTM operation, implemented as per Algorithm \ref{alg2} in Section \ref{Skyroad Operation}, facilitates real-time airspace management. The MATLAB-based implementation demonstrates the algorithm’s effectiveness through visualizations and real-time decision-making, ensuring safety, efficiency, and optimal utilization of skyroads.

\subsection{C-UTM operation Scenarios}

\begin{figure}[h]
\centering
\subfigure[]{\includegraphics[width=0.98\linewidth]{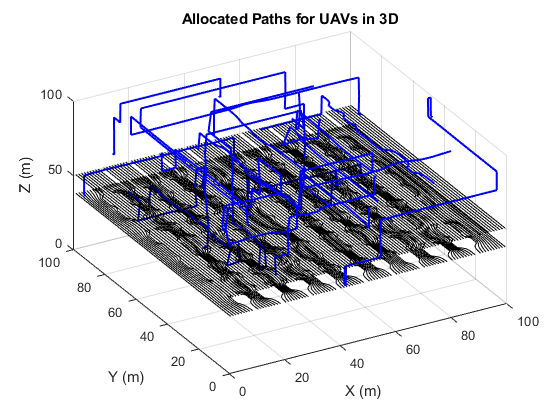}}
\subfigure[]{\includegraphics[width=0.98\linewidth]{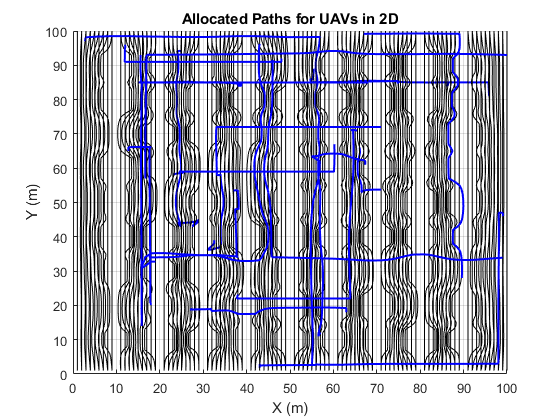}}
 \vspace{-0.5cm}
\caption{Allocated paths throughout the simulation are shown with blue solid lines in (a) three dimensions and (b) two dimensions. Only layers 3 and 4 are shown as streamline references (skyroads).}
  \label{a+b}
\end{figure}

The MATLAB simulation was executed for 500 time steps to evaluate the C-UTM framework for UASs navigating an obstacle-laden urban airspace. The allocated UAS paths throughout the simulation are illustrated in Figure \ref{a+b}, where blue solid lines represent the assigned skyroads. Figures \ref{a+b}(a) and \ref{a+b}(b) further provide three-dimensional and top-down views of the airspace, respectively, with streamlines of UAS corridors shown for layers three and four.




\begin{table*}[h!]
\centering
\renewcommand{\arraystretch}{1} 
\setlength{\tabcolsep}{15pt}    
\begin{tabularx}{0.75\textwidth}{|X|X|X|X|}
\hline \textbf{Chronological order of UASs based on airspace usage request} & \textbf{Time of new requests (in time steps), $T_{\text{request}}$} & \textbf{Maximum time needed to reach goal locations (in time steps), $T_{\text{max}}$} & \textbf{Time to check the airspace (in time steps), $T_{\text{check}}$} \\ \hline
1 & 8  & 37  & 45  \\ \hline
2 & 26  & 49  & 75 \\ \hline
3 & 31  & 101  & 132 \\ \hline
4 & 52 & 75  & 127 \\ \hline
5 & 82 & 68  & 150 \\ \hline
6 & 97  & 17  & 114  \\ \hline
7 & 145  & 51  & 196 \\ \hline
8 & 165 & 111  & 276 \\ \hline
9 & 182 & 135  & 317 \\ \hline
10 & 200 & 86  & 286 \\ \hline
11 & \cellcolor{blue!25}208  & 47  & \cellcolor{blue!25}255  \\ \hline
12 & 213  & 103  & 316 \\ \hline
13 & 238  & 138  & 376 \\ \hline
14 & 263 & 75  & 338 \\ \hline
15 & 311 & 81  & 392 \\ \hline
16 & 326  & 44  &  370 \\ \hline
17 & 355  & 48  & 403 \\ \hline
18 & 445  & 1  & 446 \\ \hline
19 & 461 & 57  & 518 \\ \hline
20 & 484 & 38  & 522 \\ \hline
\end{tabularx}
\caption{Time parameters (in time steps) throughout the simulation of a multi-agent system.} 
\label{Table}
\end{table*}

\begin{figure}[!t]
  \centering

  \hfill
  \includegraphics[width=\columnwidth,height=3.5in,keepaspectratio]{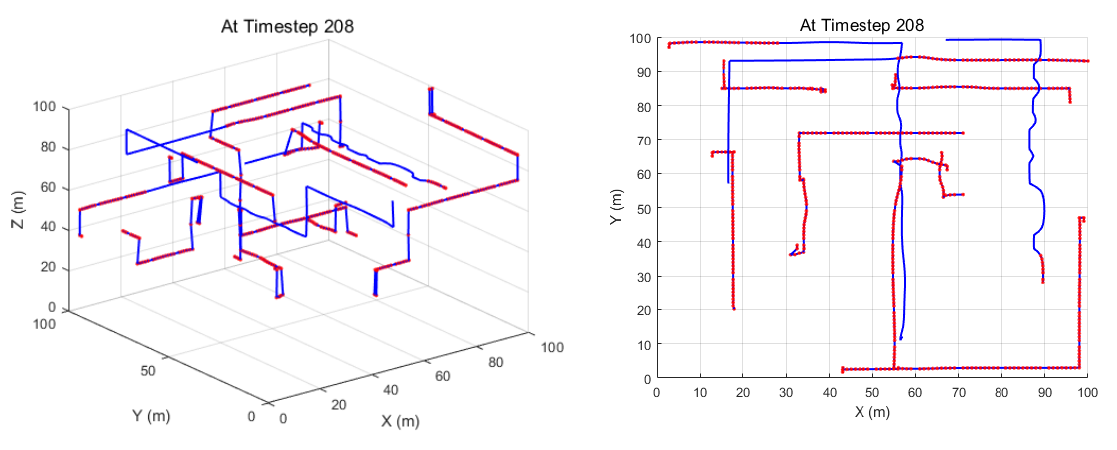}
  \caption{Left: 3D view, Right: 2D view; Allocated (blue) and traveled (red) paths at timestep 208 ($T_{\text{request}}$; Table~\ref{Table}).}
  \label{a1+b1}
\end{figure}

\begin{figure}[!t]
  \centering

  \hfill
  \includegraphics[width=\columnwidth,height=3.5in,keepaspectratio]{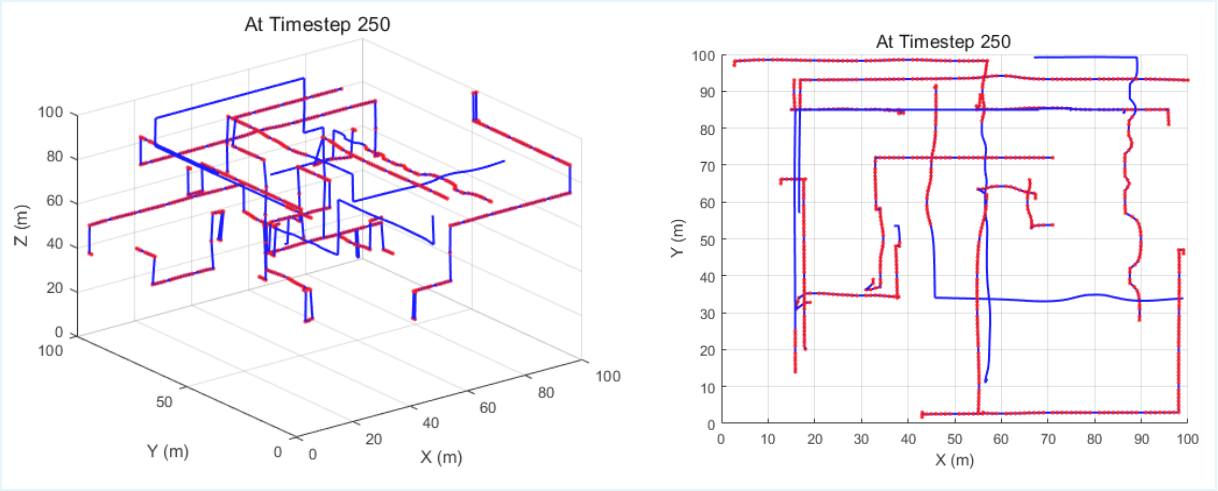}
  \caption{Left: 3D view, Right: 2D view; Allocated (blue) and traveled (red) paths at timestep 250 ($T_{\text{clear}}$; Table~\ref{Table}).}
  \label{a2+b2}
\end{figure}

\begin{figure}[!t]
  \centering

  \hfill
  \includegraphics[width=\columnwidth,height=3.5in,keepaspectratio]{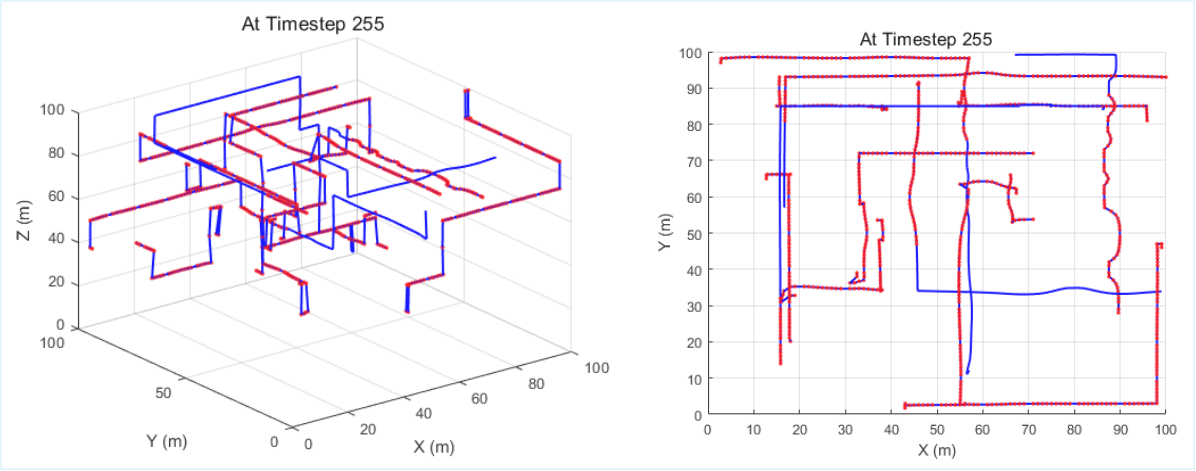}
  \caption{Left: 3D view, Right: 2D view; Allocated (blue solid) and traveled (red dotted) paths at timestep 255 ($T_{\text{check}}$; Table~\ref{Table}).}
  \label{a3+b3}
\end{figure}

Figures \ref{a1+b1}, \ref{a2+b2}, and \ref{a3+b3} demonstrate the C-UTM operation over multiple time steps, highlighting key moments ($T_{\text{request}}$, $T_{\text{clear}}$ and $T_{\text{check}}$) in UAS path allocation and traversal. All time steps that are multiples of 50 are labeled as $T_{\text{clear}}$ throughout the simulation. The $T_{\text{request}}$ and $T_{\text{check}}$ time steps are analyzed with reference to Table \ref{Table}. UAS trajectories were generated by randomly assigning start and goal locations. The framework is inherently scalable; the fleet size is parameterized by \(n\), enabling simulation of any number of vehicles with \(n \in \mathbb{N}\) and \(n \ge 1\). Without loss of generality, we evaluated a scenario with 20 UASs, each submitting an airspace-usage request randomly within 500 time steps. Each request was assigned a UAS identifier in chronological order, with varying time steps required to reach their respective destinations. For example, UAS 3 required 101 time steps, while UAS 4 reached its goal in 75 time steps (Table \ref{Table}). The length of each UAS’s assigned path was assumed to be equal to the number of time steps needed to travel from initial to final position.

As outlined in Algorithm \ref{alg2} (Section \ref{Skyroad Operation}), the airspace was cleared at two specific instances: (i) at every $T_{\text{clear}}$, occurring at time steps divisible by $N_T=50$, and (ii) at every $T_{\text{check}}$, triggered when a UAS reached its goal location. Before new path allocations at $T_{\text{request}}$, the airspace was reset to prevent previously traveled {\color{black}skyroad segments} from being treated as obstacles for incoming UAS requests. The $A^*$ search algorithm was employed for path assignment, considering occupied {\color{black}skyroad segments} from the most recent $T_{\text{clear}}$ or $T_{\text{check}}$ to ensure safe and efficient routing.

Specific scenarios can be detailed from Table \ref{Table} and Figures \ref{a1+b1}, \ref{a2+b2}, and \ref{a3+b3} for further clarification. For example, Figure \ref{a1+b1} presents the allocated (blue solid lines) and traveled (red dotted lines) paths of first ten UASs at time step 208, where UASs 1-7 have reached their goals, as confirmed by the complete overlap of red dots along their paths. At time step 208, UAS 11's request was processed using obstacle data from $T_{\text{clear}} =200$, meaning 8 occupied {\color{black}skyroad segments} from UASs 8-10 paths remained as constraints. At time step 213, UAS 12 submitted a new request. Consequently, only the 13 {\color{black}skyroad segments} occupied between steps 200 and 213 were considered as obstacles when planning UAS 12’s route. At timestep 238, UAS 13 requested to use the airspace. While assigning the optimal path for UAS 13 using A$^*$ search algorithm, traveled skyroad segments of UASs 8-12 from 201 to 238 time steps were considered as inaccessible locations. Figure \ref{a2+b2} illustrates the path allocation and path traveled at $T_{\text{clear}}=250$, when all traveled {\color{black}skyroad segments} between steps 200 and 250 were removed from the obstacle list. Figure \ref{a3+b3} presents the scenario at $T_{\text{check}}=255$, when UAS 11 just reached its goal location. UASs 8-10 and 11-13 were still en route towards their goals.

These results demonstrate the effectiveness of the C-UTM framework in dynamically managing UAS trajectories while ensuring efficient airspace utilization and real-time conflict resolution.

\subsection{C-UTM scenarios on Individual Layers}

\begin{figure*}[h!]
  \centering
  \setlength{\tabcolsep}{3pt} 
  \begin{tabular}{@{}ccc@{}}

    \includegraphics[width=\textwidth,keepaspectratio]{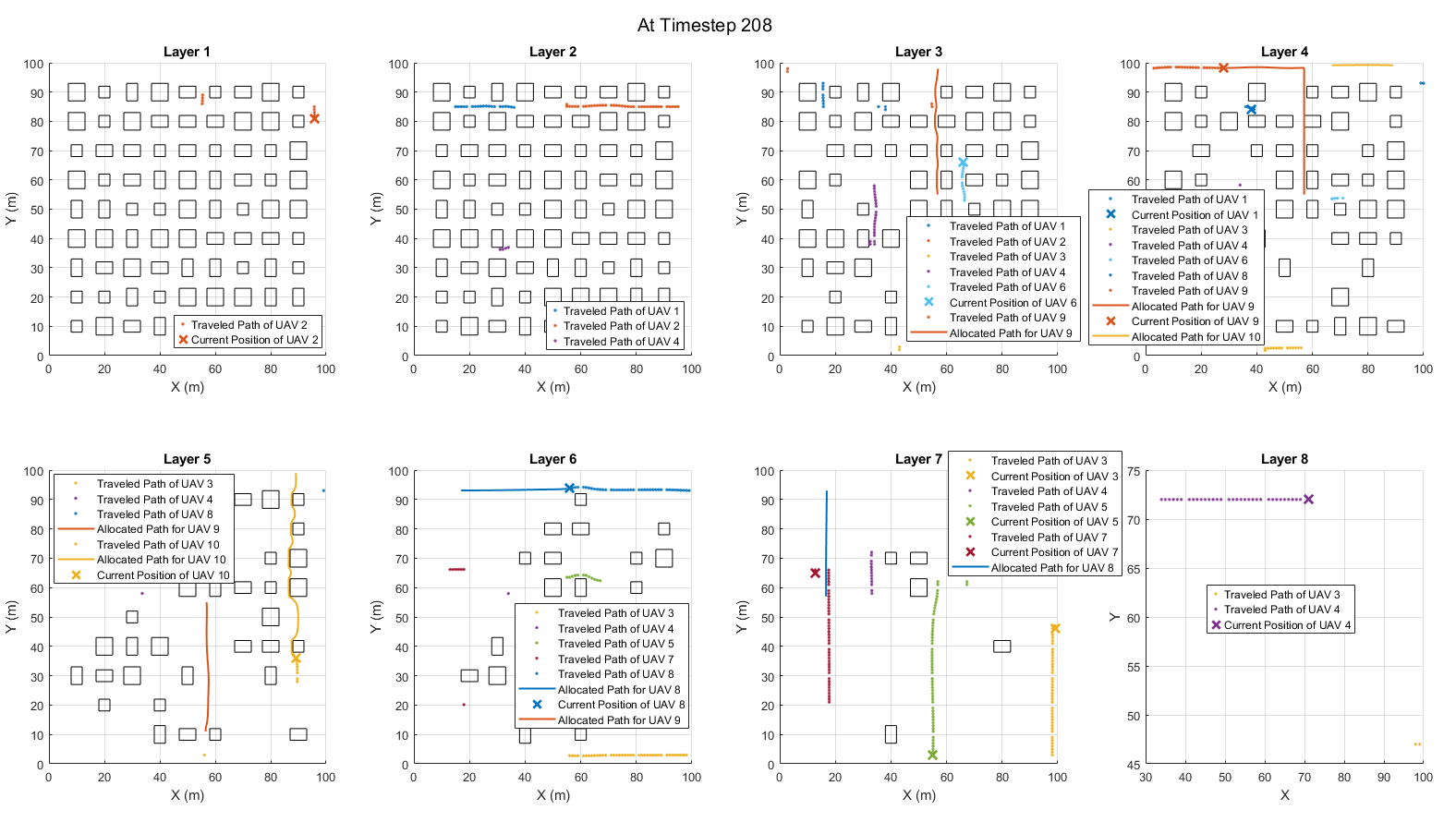} \\

  \end{tabular}
  \caption{C-UTM operation on individual layers at timestep, $T_{\text{request}} = 208$}
  \label{T_request} 
\end{figure*}

\begin{figure*}[h!]
  \centering
  \setlength{\tabcolsep}{3pt} 
  \begin{tabular}{@{}ccc@{}}
    
    \includegraphics[width=\textwidth,keepaspectratio]{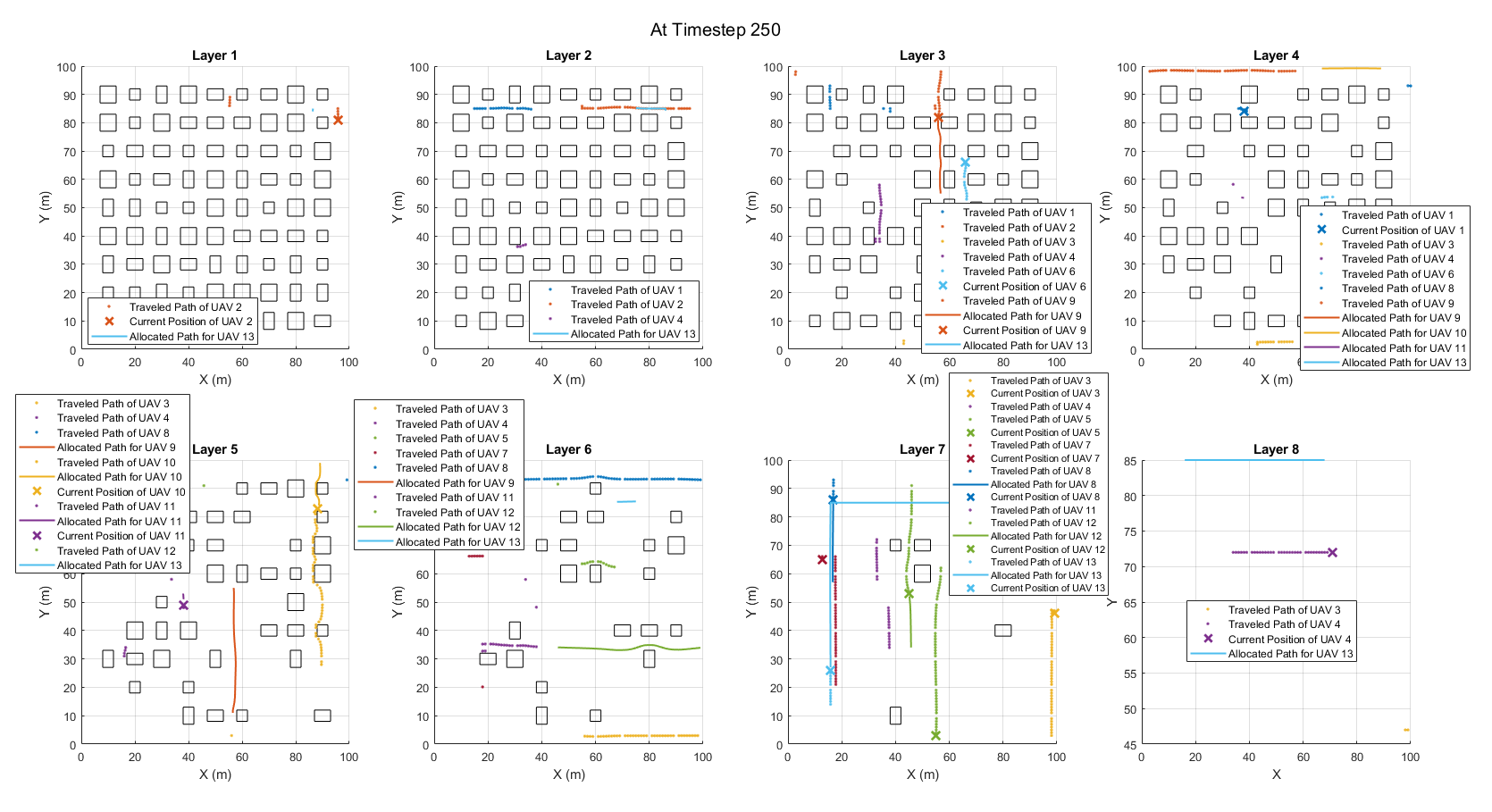}\\
    
  \end{tabular}
  \caption{C-UTM operation on individual layers at timestep, $T_{\text{clear}} = 250$}
  \label{T_clear} 
\end{figure*}

\begin{figure*}[h!]
  \centering
  \setlength{\tabcolsep}{3pt} 
  \begin{tabular}{@{}ccc@{}}
    \includegraphics[width=\textwidth,keepaspectratio]{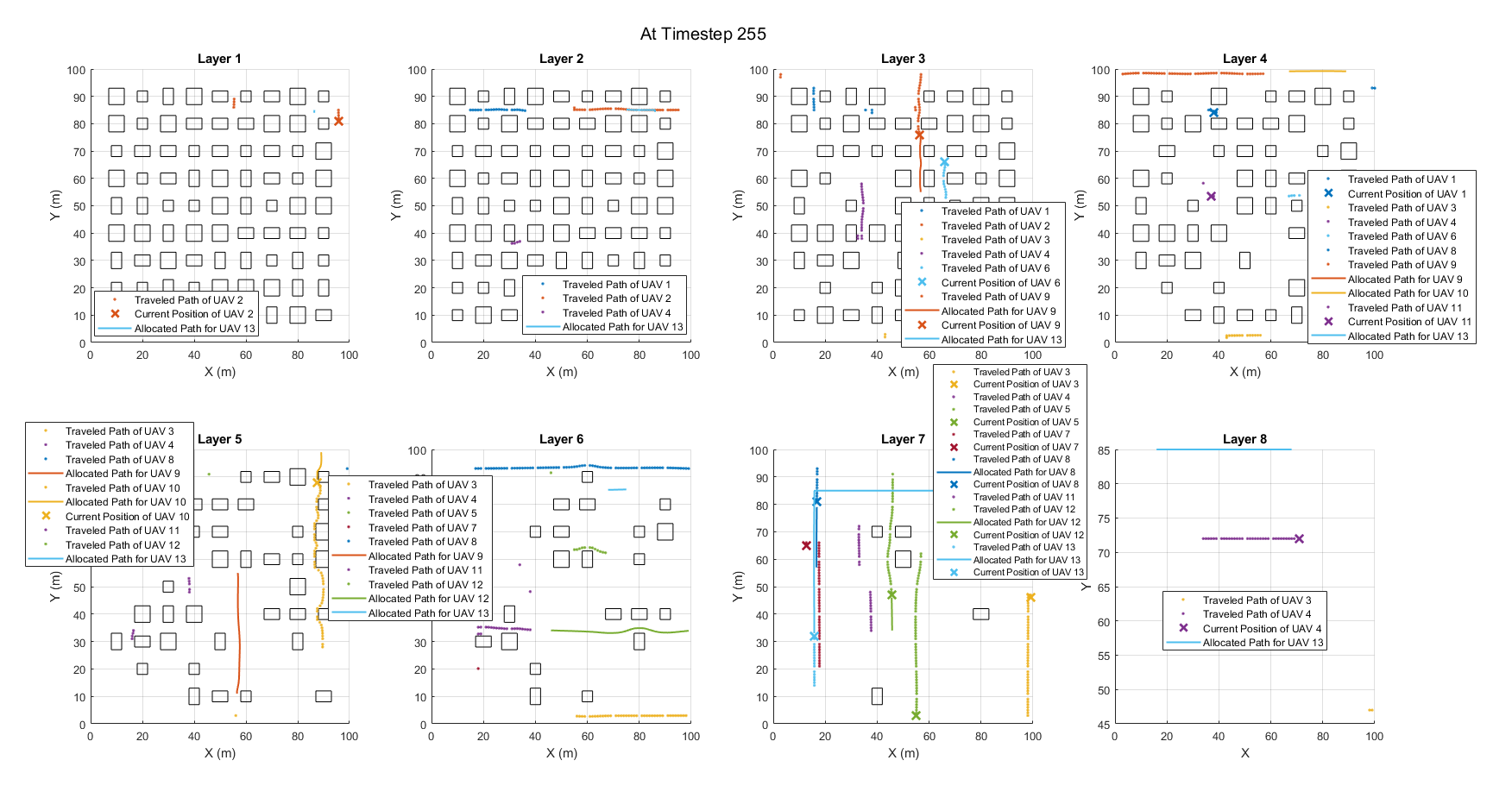} \\

  \end{tabular}
  \caption{C-UTM operation on individual layers at timestep, $T_{\text{check}} = 255$}
  \label{T_check}
\end{figure*}

To further evaluate the safety and effectiveness of the C-UTM framework and the $A^*$ search-based path planning, UAS path allocation, traveled paths, and real-time UAS positions were analyzed across each altitude layers. Figures \ref{T_request}, \ref{T_clear}, and \ref{T_check} illustrate the allocated paths (solid lines), traveled paths (dotted lines), and the current positions of UASs (marked with cross signs) at different time steps. Different colors were used to distinguish particular UAS paths clearly. 

At \(T_{\text{request}}=208\) time steps (Figure~\ref{T_request}), UASs~1--7 have reached their goal locations on distinct layers: UAS~1 on layer~4; UAS~2 on layer~1; UASs~3, 5, and 7 on layer~7; UAS~4 on layer~8; and UAS~6 on layer~3. UASs~8--10 remain en route along their allocated paths on layers~6, 4, and 5, respectively.

At \(T_{\text{clear}}=250\) (Figure~\ref{T_clear}), UASs~8--13 are in transit toward their destinations: UASs~8, 12, and 13 traverse layer~7, UAS~9 proceeds on layer~3, and UAS~10 advances on layer~5.

At \(T_{\text{check}}=255\) (Figure~\ref{T_check}), among those still in motion, UAS~11 arrives at its goal on layer~4. Consequently, all skyroad segments occupied during \(200\text{--}255\) time steps are removed from the obstacle set prior to processing the next request (UAS~14). This update ensures the airspace graph reflects only active reservations, thereby supporting timely and deconflicted path allocation for subsequent entrants.

{\color{black} Note that when goal locations lie above ground floor, vehicles can perform functions like en-route transit on skyroads, time-based sequencing or holding, rendezvous or formation operations, mission execution aloft (inspection, mapping, surveillance etc.), energy management or diverts to upper-layer chargers and so on.} 

The above results validate the effectiveness of the $A^*$ search algorithm and safety of C-UTM operation in generating optimal, obstacle-free paths for UAS navigation in each layer.

\section{Conclusion}\label{Conclusion}
The C-UTM operation successfully enables multiple UASs to operate, enter, and exit the airspace simultaneously without collisions or disturbances, ensuring efficient operation of multi-agent system. The $A^*$ algorithm ensures optimal path selection with minimal computational overhead, allowing for efficient real-time updates. The entire simulation was executed within a feasible runtime, demonstrating scalability with an increasing number of UASs. The framework effectively manages dynamic entry and exit, adapting to evolving air traffic conditions.

This simulation lays the groundwork for further research into integrating environmental uncertainties, such as dynamic weather conditions and incomplete UAS trajectory knowledge, through stochastic models like Markov Decision Processes (MDPs). Additionally, incorporating reinforcement learning techniques could enhance the system’s adaptability to complex real-world airspace scenarios.

\bibliographystyle{IEEEtran}
\bibliography{cas-refs}

 \begin{IEEEbiography}[{\includegraphics[width=1in,height=1.25in,clip,keepaspectratio]{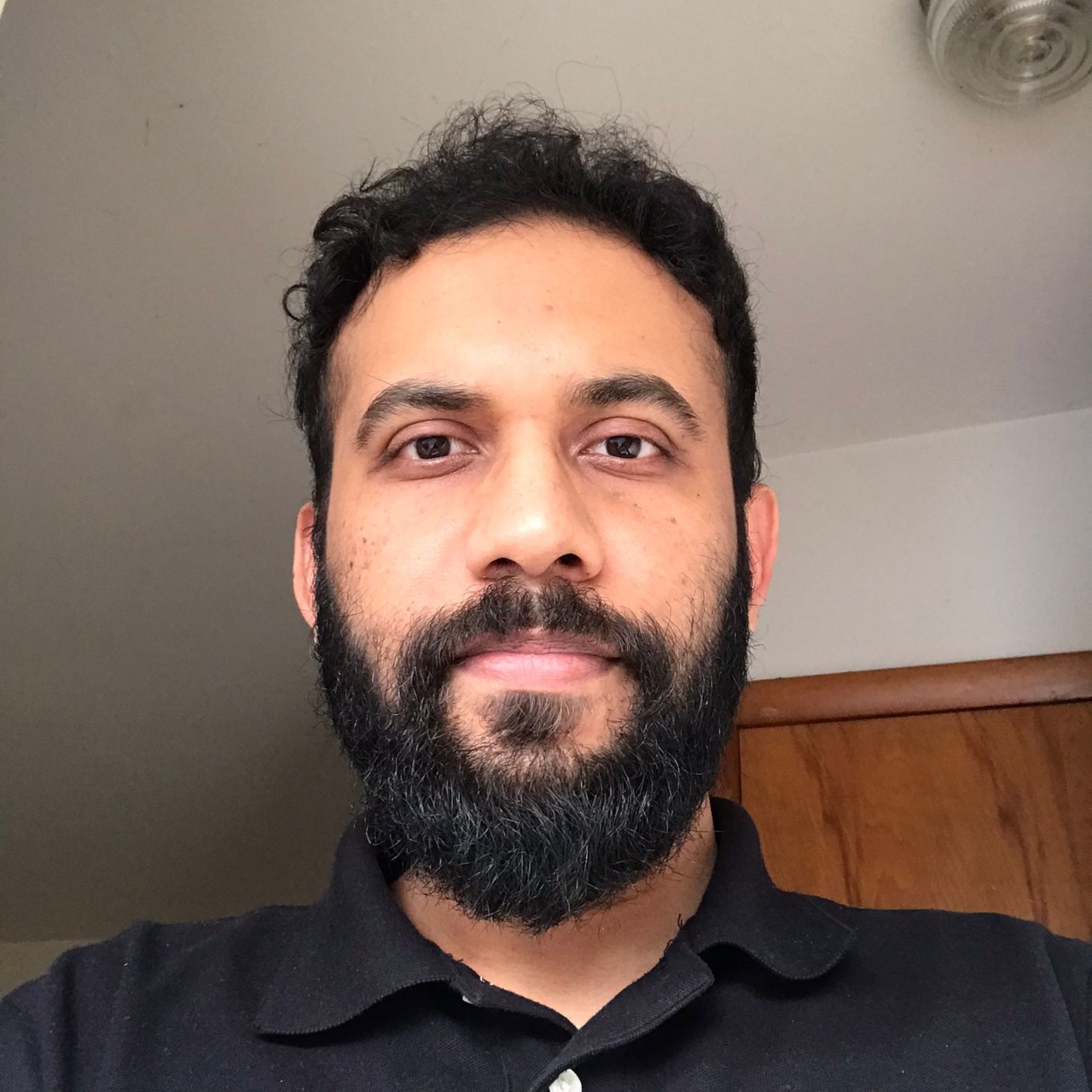}}]
{\textbf{Muhammad Junayed Hasan Zahed}} is a Graduate Research Assistant and a Ph.D. candidate in Aerospace Engineering at the University of Arizona. From 2022 to 2024, he served as a Graduate Research Assistant at The Pennsylvania State University. Previously, he was a Senior Faculty member (2020–2021) in the Department of Mechanical Engineering at the University of Creative Technology, Chittagong (UCTC). He holds a B.Sc. in Mechanical Engineering from Bangladesh University of Engineering and Technology (BUET), an M.S. in Mechanical Engineering from the University of Missouri–Kansas City (UMKC), and an M.S. in Aerospace Engineering from The Pennsylvania State University. His research interests span multi-agent systems, autonomous robots, control and developing algorithms for robotics applications. 

\end{IEEEbiography}

\begin{IEEEbiography}[{\includegraphics[width=1in,height=1.25in,clip,keepaspectratio]{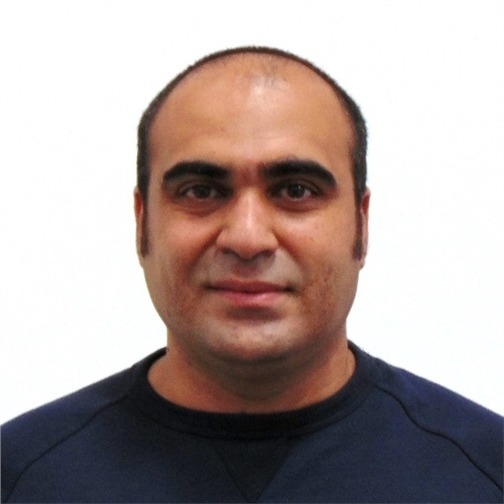}}]
{\textbf{Hossein Rastgoftar}} an Assistant Professor at the University of Arizona. Prior to this, he was an adjunct Assistant Professor at the University of Michigan from 2020 to 2021. He was also an Assistant Research Scientist (2017 to 2020) and a Postdoctoral Researcher (2015 to 2017) in the Aerospace Engineering Department at the University of Michigan Ann Arbor. He received the B.Sc. degree in mechanical engineering-thermo-fluids from Shiraz University, Shiraz, Iran, the M.S. degrees in mechanical systems and solid mechanics from Shiraz University and the University of Central Florida, Orlando, FL, USA, and the Ph.D. degree in mechanical engineering from Drexel University, Philadelphia, in 2015. 
\end{IEEEbiography}

\vfill

\end{document}